\documentclass[aps
        ,pra
        ,draft
        ,showpacs
	,showkeys
]{revtex4}

\begin{document}

\title{Entanglement correlations, Bell inequalities and the
concurrence}
\author{Markus A. Cirone}
\affiliation{ECT*, Strada delle Tabarelle 286, I-38050 Villazzano, Trento, Italy}


\begin{abstract}
Probabilities of measurement outcomes of two-particle entangled states 
give a physically transparent interpretation of the concurrence and of the
I-concurrence as entanglement measures. The (I-)concurrence can thus
be measured experimentally. The tight connection between
these measures and Bell inequalities is highlighted.
\end{abstract}

\pacs{03.65.Ud, 03.67.-a}
\keywords{Entanglement measure;  concurrence; Bell inequalities}

\maketitle

\section{Introduction}

Entanglement is the most puzzling feature of quantum mechanics.
The surprising behaviour of entangled states has been stressed by
Einstein, Podolsky, and Rosen (EPR) \cite{epr}, who considered
a two--particle system in an entangled state and showed that
the assumptions of locality, realism and completeness of
quantum mechanics fail to give a consistent physical description
of quantum systems. Several years later, Bell found his celebrated
inequalities \cite{bell} which show that entanglement cannot
be consistent with any local theory containing hidden variables.
Many--particle entanglement has been investigated much later \cite{drum}.
The best known example of multipartite entanglement are probably
the GHZ states of three particles \cite{ghz1,ghz2}.

In the latest years investigation of entanglement has received
new impulse by the development of quantum information theory \cite{nich}.
In this new field entangled particles are
more powerful a resource than separable ones in a number of
possible applications, ranging from teleportation
\cite{tel1,tel2,tel3,tel4,tel5,tel6,tel7,tel8}
to investigations of quantum channels \cite{us,palm}. Therefore
the question ``How much are two or more particles entangled?''
has been raised, since particles with a high degree of
entanglement should be a better resource than less entangled ones.
This question has led to the definition of several measures of
entanglement \cite{meas}.

It is a remarkable fact that Bell's approach to the mysteries of
entanglement has played so far a marginal role, if any, in the definition
of the entanglement measures. It often occurs that
the investigations focus on the fulfillment of the formal properties
of entanglement measures \cite{plen}, sometimes to the detriment of a clear
and deeper physical insight comparable to that given by Bell inequalities.
A physical interpretation of entanglement measures is however desirable,
since it helps us in understanding the essence of entanglement and
can indicate new directions for further studies on this subject.
A connection between Bell's ideas and entanglement measures would be
a step forward in this direction.

The present paper fills the gap between Bell inequalities
and entanglement measures. The fundamental hypothesis in the derivation of
Bell inequalities is the locality assumption. Locality leads to statistical properties
that characterize the joint and local probabilities for outcomes of measurements
on the two particles and do not hold for entangled pure states.
We shall use these probabilities to define an entanglement measure for pure states,
that turns out to be equal to the (squared) concurrence introduced by
Hill and Wootters \cite{hill,woo1,woo2} and to the Renyi 2-entropy \cite{reny,horo},
also known as linear entropy \cite{breu} or I-concurrence \cite{rung}.
Our approach is completely independent of the motivations that led
to their definition and prompts a transparent physical interpretation for them.
Moreover, it prompts a simple way to measure the concurrence experimentally.
In order to illustrate our idea, we first examine pure states.
Only at the end of the paper we shall extend our considerations to mixed states, where the
same interpretation is shown to hold for a particular decomposition of the
density matrix obtained via a minimization procedure, as in \cite{woo1}.

The paper is organized as follows: In Section II we briefly review Bell's argument
for two two-level entangled pure states. We shall emphasize the importance of locality
in Bell's derivation of his inequalities and then show how locality and nonlocality
affect joint and conditional probabilities of measurement outcomes. These considerations,
though elementary, have never appeared in articles or textbooks discussing entanglement,
to the best of our knowledge. In Section III we shall then show that for pure states these probabilities
lead to a transparent alternative interpretation of the (I-) concurrence that
stresses its intimate relation with Bell inequalities. The result, valid for
pure states with Hilbert spaces of arbitrary dimensions, is then generalized to mixed states and sheds light
also on three-partite entanglement. Section IV contains the conclusions.

\section{EPR correlations of pure states}

Bell's analysis of the EPR problem considers two two-level systems, like, e.g.,
two linearly polarized photons in the pure entangled state

\begin{equation}
\mid \psi_{{\rm EPR}} \rangle = \frac{1}{\sqrt{2}} (\mid V_{{\rm A}}\rangle \mid V_{{\rm B}}\rangle-
\mid H_{{\rm A}}\rangle \mid H_{{\rm B}} \rangle)
\label{entb}
\end{equation}
where $V,H$ refer respectively to the vertical and horizontal polarization
of the two photons along some given directions (for more details on Bell's analysis, see
Refs. \cite{bell,ghz1,ghz2,lalo}).
We can assume that the two photons, denoted as $A$ and $B$, are in two different and distant
laboratories where two physicists, usually called Alice and Bob,
perform polarization measurements.

Bell examines the statistics of measurements on the two particles; in particular
he argues that the assumption of locality leads to write the expectation value
of a joint measurement of polarizations along two directions $\hat{n}_A$ and $\hat{n}_B$ as

\begin{equation}
E(\hat{n}_A,\hat{n}_B)=\int A_{\lambda}(\hat{n}_A)B_{\lambda}(\hat{n}_B)d\rho
\label{expe}
\end{equation}
where $\lambda$ represents the set of hidden variables, $\rho$ is the probability measure
and the functions $A_{\lambda}(\hat{n}_A)$ and $B_{\lambda}(\hat{n}_B)$ are the outcomes
of measurements on photons $A$ and $B$, respectively (their values can be set to
$\pm 1$ for $V$ and $H$ polarizations, respectively). Bell shows then that expectation values like
Eq.(\ref{expe}) have to satisfy some inequalities, that are however violated by entangled states
like Eq.(\ref{entb}). The assumption of locality appears through the factorization
$A_{\lambda}(\hat{n}_A)B_{\lambda}(\hat{n}_B)$ in Eq.(\ref{expe}) and the absence of $\hat{n}_B$ ($\hat{n}_A$)
in the function $A_{\lambda}(\hat{n}_A)$ ($B_{\lambda}(\hat{n}_B)$).

Along the same line of thought, we investigate how nonlocal correlations manifest
themselves in the wavefunction, from which all statistical properties of the system are derived.
In particular we examine the probabilities of outcomes of polarization measurements performed
on the two particles. We stress that our considerations hold for pure states. Another important
point is that we shall always use the Schmidt decomposition \cite{knek} and consider only measurements
that project the wavefunction on one of the states of the Schmidt bases.

Let's examine probabilities of outcomes of polarization measurements for a
general two-photon entangled state, having Schmidt decomposition

\begin{equation}
\mid \psi_{\rm AB} \rangle_{\rm E} \equiv \sqrt{\lambda}\mid V_{\rm A} \rangle \mid V_{\rm B} \rangle
+\sqrt{1-\lambda} \mid H_{\rm A} \rangle \mid H_{\rm B} \rangle
\label{schm}
\end{equation}
In this equation V (H) denotes vertical (horizontal) polarization along
the axes that correspond to the polarization states
of the Schmidt bases. These directions may be different for the two photons. The
parameter $0 < \lambda < 1$ is the Schmidt parameter.
Photons are found with vertical or horizontal polarizations with local probabilities

\begin{equation}
P(V_{\rm A})=P(V_{\rm B})=\lambda, \;\;\;\;\;\;\; P(H_{\rm A})=P(H_{\rm B})=1-\lambda
\label{prob}
\end{equation}
respectively. From these results Alice and Bob cannot conclude that the photon pair is
in an entangled state, because the pure, separable state

\begin{equation}
\mid \psi_{\rm AB} \rangle_{\rm S} \equiv ( \sqrt{\lambda}\mid V_{\rm A}\rangle +
\sqrt{1-\lambda}\mid H_{\rm A}\rangle) \otimes
(\sqrt{\lambda}\mid V_{\rm B}\rangle + \sqrt{1-\lambda}\mid H_{\rm B} \rangle)
\label{sepb}
\end{equation}
gives the same local probabilities.
When they compare the results of their measurements, however,
they find that both photons have either vertical or
horizontal polarization. These correlations do not
occur when the two photons are in the separable state Eq.(\ref{sepb}).

Entanglement manifests itself in the joint probabilities
and in the conditional probabilities of measurement outcomes.
For the experiment here described, there are four different
pairs of measurement outcomes: $V_{A}V_{B}$,
$V_{A}H_{B}$, $H_{A}V_{B}$, and $H_{A}H_{B}$, with joint probabilities
$P(V_{A},V_{B})$, $P(V_{A},H_{B})$, $P(H_{A},V_{B})$, and
$P(H_{A},H_{B})$, respectively, to occur.
If the two photons are in the separable state Eq.(\ref{sepb}),
the measurements outcomes are
independent and the joint probabilities factorize accordingly

\begin{equation}
P_{\rm S}(X_{\rm A},Y_{\rm B}) = P(X_{\rm A})P(Y_{\rm B})
\label{fact}
\end{equation}
into the product of the local probabilities $P(X)$ and $P(Y)$
given by Eq.(\ref{prob}) ($X,Y=H,V$).
For the entangled state Eq.(\ref{schm}) the factorization does not hold,

\begin{equation}
P_{\rm E}(X_{\rm A},Y_{\rm B}) \neq P(X_{\rm A})P(Y_{\rm B})
\label{nonf}
\end{equation}
An equivalent description makes use of the conditional probabilities

\begin{equation}
P(X_{\rm A} \mid Y_{\rm B})=\frac{P(X_{\rm A},Y_{\rm B})}{P(Y_{\rm B})}
\end{equation}
of finding photon A with polarization $X$ after photon B has been
found with polarization $Y$. Indeed, for the separable state
Eq.(\ref{sepb}) we find

\begin{equation}
P_{\rm S}(X_{\rm A} \mid Y_{\rm B}) = P(X_{\rm A})
\label{cond}
\end{equation}
so the conditional probabilities are equal to the local probabilities
(in other words, measurements on B do not condition the results of
measurements on A), whereas for the entangled state Eq.(\ref{schm})
similar equalities do not hold:

\begin{equation}
P_{\rm E}(X_{\rm A} \mid Y_{\rm B}) \neq P(X_{\rm A})
\label{con1}
\end{equation}
For entangled states measurements on one of the particles modifies
the probabilities of outcomes of measurements on the second particle.
These considerations follow closely Bell's approach of the EPR problem.
Incidentally, we note that factorization of joint probabilities
can be an easier check for entanglement than violation of Bell inequalities.

We conclude this section by stressing the importance of using
the Schmidt decomposition in the arguments illustrated above.
In fact, it may occur that for other bases the properties
of the conditional and joint probabilities are different.
For instance, bases exist such that some conditional
probabilities are equal to local probabilities even for
entangled pure states. The choice of the Schmidt basis helps us to express probabilities
in terms of invariant parameters, an important requisite for the definition
of an entanglement measure.

\section{The entanglement measure through probabilities}

In the previous section we have seen that entanglement of pure states is strictly
related to the occurrence of correlations in the measurement outcomes, as
pointed out by Bell. We have also seen how these correlations affect joint and
conditional probabilities. It is thus natural to wonder if these probabilities can be
used to define a measure of entanglement, in particular in the case
of pure states, where correlations are a clear sign of entanglement.
In fact, it is possible to use the local, joint and
conditional probabilities for this purpose. We first illustrate
the idea using the polarization states
Eqs.(\ref{schm}) and (\ref{sepb}) and then we
generalize our approach to bipartite states of Hilbert spaces
of arbitrary dimensions.

For polarization measurements on the Schmidt bases the inequalities
Eqs.(\ref{nonf}), (\ref{con1}), are due to correlations.
These considerations on the conditional and joint probabilities
are not independent, but can rather be merged together
when we define the quantity

\begin{eqnarray}
\Delta(X_{\rm A},Y_{\rm B})\equiv
\mid P(X_{\rm A} \mid Y_{\rm B})-P(X_{\rm A}) \mid P(Y_{\rm B})
\nonumber \\
=\mid P(X_{\rm A},Y_{\rm B})-P(X_{\rm A})P(Y_{\rm B}) \mid
\label{diff}
\end{eqnarray}
that points out the entanglement correlations between measurements
on the two particles, since it vanishes in the absence
of correlations. The quantity $\Delta(X_{\rm A},Y_{\rm B})$
can be interpreted as the difference of conditional or joint
probabilities of the entangled state Eq.(\ref{schm})
and the corresponding separable state
(\ref{sepb}) that gives the same local probabilities.
In this sense, $\Delta(X_{A},Y_{B})$ measures the deviation
from separability for two particular measurements. We define then the measure of entanglement of two photons
in an arbitrary polarization state $\mid \psi_{\rm AB} \rangle$
as the sum of $\Delta$'s for all four possible combinations of polarizations,

\begin{equation}
E_{2}(\psi_{\rm AB})\equiv \Delta(H_{\rm A},H_{\rm B})+\Delta(H_{\rm A},V_{\rm B})+
\Delta(V_{\rm A},H_{\rm B})+\Delta(V_{\rm A},V_{\rm B})
\label{e2}
\end{equation}
which for state (\ref{schm}) gives

\begin{equation}
E_{2}(\psi_{\rm AB})=4\lambda(1-\lambda)
\label{e2bis}
\end{equation}
i.e., the (squared) concurrence defined by Hill and Wootters \cite{hill,woo1,woo2}. We point out that
the concurrence can thus be measured, or, more precisely, estimated experimentally if the Schmidt decomposition
is known, in the same way as violations of Bell inequalities are verified.

Now we generalize the definition of entanglement measure to Hilbert spaces of arbitrary
dimensions $M$ and $M'$, respectively. Since different generalizations of the concurrence
have been proposed \cite{rung,othe}, it is interesting to see if our generalization leads
to any of them. Any state of the two particles has a Schmidt decomposition of the form

\begin{equation}
\mid \psi_{\rm AB} \rangle = \sum_{i=1}^{N} \sqrt{\lambda_{i}}
\mid i_{\rm A} \rangle \otimes \mid i_{\rm B} \rangle,
\label{schmn}
\end{equation}
where $N={\rm min}(M,M')$. We consider again the local
probabilities $P(i_{\rm A})$, ($P(j_{\rm B})$) of finding particle A
(B) in the state $\mid i_{\rm A} \rangle$ ($\mid j_{\rm B} \rangle$)
of the Schmidt basis and, accordingly, the joint probabilities
$P(i_{\rm A},j_{\rm B})$ or the conditional probabilities
$P(i_{\rm A} \mid j_{\rm B})$, that for the state Eq.(\ref{schmn}) are

\begin{eqnarray}
P(i_{\rm A})=P(i_{\rm B})=\lambda_i, \;\;\;\;\; P(i_{\rm A},j_{\rm B})=\delta_{ij}\lambda_{j},
\;\;\;\;\; & P(i_{\rm A} \mid j_{\rm B} )=\delta_{ij}, &
\label{pron}
\end{eqnarray}
(here $\delta_{ij}$ represents Kronecker's delta),
and define the quantity

\begin{equation}
\Delta(i_{\rm A},j_{\rm B}) \equiv
\mid P(i_{\rm A},j_{\rm B})-P(i_{\rm A})P(j_{\rm B}) \mid
=\mid P(i_{\rm A} \mid j_{\rm B})-P(i_{\rm A})\mid P(j_{\rm B})
\end{equation}
On the analogy of Eq.(\ref{e2}), the entanglement measure for
arbitrary dimensions of Hilbert spaces is defined as

\begin{eqnarray}
E_{N}(\psi_{\rm AB}) & \equiv & \frac{N}{2(N-1)} \sum_{i,j=1}^{N}
\Delta(i_{\rm A},j_{\rm B})
\label{en}
\end{eqnarray}
where the normalization factor $N/[2(N-1)]$ has been introduced in order to ensure
$0\le E_N \le 1$. With the expressions (\ref{pron}) obtained for the probabilities the
entanglement measure becomes

\begin{eqnarray}
E_N(\psi_{\rm AB}) & = & \frac{N}{2(N-1)} \sum_{i,j=1}^{N}
\mid P(i_{\rm A} \mid j_{\rm B})-P(i_{\rm A})\mid P(j_{\rm B})
\nonumber \\
& = & \frac{N}{2(N-1)} \sum_{i=1}^{N}\sum_{j=1}^{N}
\mid \delta_{ij}-\lambda_{i} \mid \lambda_{j}=
\frac{N}{2(N-1)} \left[ \sum_{i=1}^{N}(1-\lambda_{i})\lambda_{i}
+\sum_{i=1}^{N}\sum_{j\neq i}\lambda_{i}\lambda_{j} \right]
\nonumber \\
& = & \frac{N}{2(N-1)} \left[ \sum_{i=1}^{N} \lambda_{i}
-\sum_{i=1}^{N} \lambda_{i}^2 +\sum_{i=1}^{N} \lambda_{i}
\left( \sum_{j=1}^N \lambda_{j}-\lambda_{i}\right) \right]
\end{eqnarray}
When the normalization condition $\sum_{i=1}^N \lambda_{i}=1$
of the state Eq.(\ref{schmn}) is taken into account,
we obtain

\begin{equation}
E_N(\psi_{\rm AB})=\frac{N}{(N-1)}
\left[ 1-\sum_{i=1}^{N} \lambda_{i}^2\right]
\label{enbis}
\end{equation}
which, apart from a numerical factor, is the Renyi 2-entropy
or I-concurrence. We have thus given an alternative derivation
of this entanglement measure that stresses its relation with correlations
and thus with Bell inequalities. We also wish to emphasize that our
approach prompts an immediate generalization of the concurrence without any further
hypothesis, contrarily to the more complex mathematical analyses carried
out in Refs. \cite{rung,othe}.

So far we have considered only pure states. The extension of our results
to mixed states proceeds along the same line of reasoning that
extends the (I-)concurrence to mixed states \cite{woo1,rung}. Among the sets of
states $\{ \mid \psi_i \rangle \}$ that decompose the density matrix $\rho_{AB}$
into a mixture of pure states

\begin{equation}
\rho_{\rm AB}=\sum_i p_i \mid \psi_i \rangle \langle \psi_i \mid
\end{equation}
we pick out the one that minimizes the average concurrence. The (I-)concurrence
is then related to joint probabilities of measurement outcomes performed on the
Schmidt basis of each state $\mid \psi_i \rangle$ of that particular decomposition.
Again, the amount of entanglement can be measured experimentally if measurements
are done on the ensemble of systems corresponding to that minimal decomposition.

To conclude, we note that the residual (3-) tangle \cite{coff} that quantifies
the essential three-way entanglement in three-qubit systems, being expressed in
terms of the squared concurrence of subsystems, is also amenable of an interpretation
based on joint probabilities.

\section{Summary and conclusions}

Our results can be summarized as follows: (i) nonlocality of entangled pure
states, that lies at the heart of violation of Bell inequalities, manifests itself in
joint and conditional probabilities of measurement outcomes, which prompt
a physical interpretation of the entanglement measure known as concurrence,
and the interpretation also holds for mixed states; (ii) being expressed in terms of probabilities of measurement outcomes,
the concurrence can be experimentally measured; (iii) our derivation does not assume concurrence, is complementary
to the more mathematically oriented definition given by Hill and Wootters
and gives physical motivations to prefer it when compared to other entanglement
measures for qubits; (iv) if non-local correlations are the essence of entanglement, then the
I-concurrence is the most natural extension of the concurrence to Hilbert spaces
of higher dimensions;
(v) in three-partite states the non local correlations that are present in joint
probabilities also give a physical interpretation of the residual 3-tangle. 
It is our hope that similar analyses can help to achieve further insight in
multiparticle entanglement.

\section{Acknowledgments}

I wish to thank I. Cirac, P. D. Drummond,
M. Freyberger, B. Kraus, M. Ku\'s, M. Lewenstein,
G. M. Palma, F. Persico, A. Sanpera and K \.Zyczkowski for
discussions about entanglement, interesting comments and
for drawing some relevant articles to my attention.
Part of this work was done when I was postdoc fellow at the
Abteilung f\"ur Quantenphysik of the University of Ulm, Germany,
supported by the IHP program of the European Union with the network `QUEST'.


\begin{thebibliography}{99}
\bibitem{epr} A. Einstein, B. Podolsky and N. Rosen,
 Phys. Rev. 47 (1935) 777.
\bibitem{bell} J. S. Bell, Physics 1 (Long Island, NY),
(1964) 195; reprinted in J. S. Bell,
Speakable and Unspeakable in Quantum Mechanics,
Cambridge University Press, Cambridge (1987).
\bibitem{drum} To the best of our knowledge, the first
investigation about Bell's inequalities in multipartite
states is in P. D. Drummond,
Phys. Rev. Lett. 50 (1983) 1407.
\bibitem{ghz1} D. M. Greenberger, M. Horne and A. Zeilinger,
in Bell's Theorem, Quantum Theory and Conceptions
of the Universe, edited by M. Kafatos,
Kluwer Academic, Dordrecht (1989) p.73.
\bibitem{ghz2} D. M. Greenberger, M. A. Horne, A. Shimony and
A. Zeilinger, Am. J. Phys. 58 (1990) 1131.
\bibitem{nich} M. A. Nielsen and I. L. Chuang, Quantum
Computation and Quantum Information, Cambridge University Press, Cambridge (2000).
\bibitem{tel1} C. H. Bennett et al., Phys. Rev. Lett. 70, (1993)  1895.
\bibitem{tel2} L. Davidovich, N. Zagury, M. Brune, J.M. Raimond,S. Haroche, Phys. Rev. A 50 (1994) R895.
\bibitem{tel3} D. Boschi, S. Branca, F. DeMartini, L. Hardy, S. Popescu, Phys. Rev. Lett. 80
 (1998) 1121.
\bibitem{tel4} D. Bouwmeester et al., Nature 390 (1997) 575.
\bibitem{tel5} A. Furusawa et al., Science 282 (1998) 706.
\bibitem{tel6} M. A. Nielsen, E. Knill and R. Laflamme, Nature 396 (1998) 52.
\bibitem{tel7} M. Riebe {\em et al}., Nature 429 (2004) 734.
\bibitem{tel8} M. D. Barrett {\em et al}., Nature 429 (2004) 737.
\bibitem{us} D. G. Fischer, H. Mack, M. A. Cirone
and M. Freyberger, Phys. Rev. A 64 (2001) 022309.
\bibitem{palm} C. Macchiavello and G. M. Palma, Phys. Rev. A 65 (2001) 050301(R).
\bibitem{meas} M. Horodecki, Quant. Inf. Comp. 1 (2001) 3.
\bibitem{plen} V. Vedral and M. B. Plenio, Phys. Rev. A 57 (1998) 1619.
\bibitem{hill} S. Hill and W. K. Wootters. Phys. Rev. Lett. 78 (1997) 5022.
\bibitem{woo1} W. K. Wootters, Phys. Rev. Lett. 80 (1998) 2245.
\bibitem{woo2} W. K. Wootters, Quant. Inf. Comp. 1 (2001) 27.
\bibitem{reny} A. R\'enyi, Probability Theory, North--Holland, Amsterdam (1970).
\bibitem{horo} R. Horodecki and M. Horodecki, Phys. Rev. A 54 (1996) 1838.
\bibitem{breu} H.-P. Breuer and F. Petruccione, The theory of open quantum systems,
Oxford University Press, Oxford (2002).
\bibitem{rung} P. Rungta, V. Buzek, C. M. Caves, M. Hillery, G. J. Milburn, Phys. Rev. A 64 (2001) 042315.
\bibitem{lalo} F. Lalo\"e, Am. J. Phys. 69 (2001) 655.
\bibitem{knek} A. Ekert and P. L.  Knight, Am. J. Phys. 63 (1995) 415.
\bibitem{othe} A. Uhlmann, Phys. Rev. A 62 (2000) 032307.
\bibitem{coff} V. Coffman, J. Kundu, and W. K. Wootters, Phys. Rev. A 61 (2000) 052306.
\end{thebibliography}
\end{document}